\documentclass[aps,prx,amsfonts,amssymb,amsmath,
  twocolumn,
   reprint,
   superscriptaddress,
   noshowpacs,
   preprintnumbers,
   amsmath,amssymb
]{revtex4-1}

\pdfoutput=1
\usepackage{graphicx}
\usepackage{dcolumn}

\usepackage{color}
\usepackage{graphicx}
\usepackage[pdftex,colorlinks=true,linkcolor=red,citecolor=blue]{hyperref}
\begin{document}

%\linenumbers

%\preprint{Applied Physics Letters; ver.1a}

\title{Thermoelectric power as a probe of density of states in correlated actinide materials: the case of PuCoGa$_{5}$ superconductor}

\author{K.~Gofryk}
\email{krzysztof.gofryk@inl.gov}
\affiliation{Idaho National Laboratory, Idaho Falls, Idaho 83415, USA}

\author{J-C.~Griveau}
\affiliation{European Commission, Joint Research Centre, Institute for Transuranium Elements, Postfach 2340, 76125 Karlsruhe, Germany}

%\author{J. Rebizant}
%\affiliation{European Commission, Joint Research Centre, Institute for Transuranium Elements, Postfach 2340, 76125 Karlsruhe, Germany}

\author{P.~S.~Riseborough}
\affiliation{Department of Physics, Temple University, Philadelphia, Pennsylvania 19122, USA}

\author{T.~Durakiewicz}
\affiliation{Los Alamos National Laboratory, Los Alamos, New Mexico 87545, USA}

\date{\today}% It is always \today, today, %  but any date may be explicitly specified

\begin{abstract}

We present measurements of the thermoelectric power of the plutonium-based unconventional superconductor PuCoGa$_{5}$. The data is interpreted within a phenomenological model for the quasiparticle density of states of intermediate valence systems and the results are compared with results obtained from photoemission spectroscopy. The results are consistent with intermediate valence nature of 5$f$-electrons, furthermore, we propose that measurements of the Seebeck coefficient can be used as a probe of density of states in this material, thereby providing a link between transport measurements and photoemission in strongly correlated materials. We discuss these results and their implications for the electronic structure determination of other strongly correlated systems, especially actinide materials.

\end{abstract}

%\pacs{}% PACS, the Physics and Astronomy
\maketitle

%\section{Introduction}

Nuclear materials are complex multi-component systems which contain actinide elements, characterized by the presence of 5$f$-electrons. In order to advance the fundamental understanding of these materials and their thermophysical properties, so important in nuclear energy production, all aspects of the so-called 5$f$-electron challenge have to be addressed.\cite{LANL,kevin} Actinides are characterized by the coexistence of localized and itinerant (delocalized) 5$f$-electrons near the Fermi energy. This dual nature of the 5$f$-electrons leads to many exotic phenomena, spanning complex magnetic ordering, heavy-fermion ground state, and/or "non-Fermi liquid" state (see Refs. \onlinecite{HvL,U122,G1}). Surprisingly, unconventional superconductivity has also been found in this class of materials. Prime examples are given by UBe$_{13}$, NpPd$_{5}$Al$_{2}$ and PuCoGa$_{5}$ (see Refs. \onlinecite{UBe13,Np152,Pu115}). All of these findings have stimulated great interest in actinide materials. However, despite  intensive theoretical and experimental efforts their electronic structures are still not well understood.\cite{el_str,gabi} The band structure of solids is usually investigated by various photoelectron spectroscopy methods. In nuclear materials, however, these experiments are very demanding due to the presence of radioactivity and radio-toxicity of actinide elements, especially plutonium. All of these require a special approach and limit the research to a few laboratories over the world.

Here we present the thermoelectric properties of the plutonium-based unconventional superconductor PuCoGa$_{5}$. It has been characterized as a strongly correlated compound with a superconducting transition temperature of $T_{c}$~=~18.5~K.\cite{Pu115,curro,jcg} After a decade of investigation, the mechanism responsible for Cooper pair formation remains unknown, as does the role of 5$f$-electrons. Photoemission measurements and electronic calculations made on PuCoGa$_{5}$ point to a strongly renormalized electronic structure with a sharp peak just below the Fermi level.\cite{joyce,jxz} We show that Seebeck coefficient of PuCoGa$_{5}$ is strongly enhanced and its magnitude and temperature dependence is characteristic of intermediate valence systems, indicating the presence of the 5$f$-band in close proximity to the Fermi level. The thermoelectric data can be well described by a phenomenological picture of strongly renormalized quasiparticle bands\cite{gott}. The results obtained point to the itinerant nature of the 5$f$-electrons in PuCoGa$_{5}$ and their importance for unconventional superconductivity. The electronic structure parameters for PuCoGa$_{5}$ obtained from the thermoelectric measurements are shown to be related to those obtained from photoemission and this relation can be extended to other valence fluctuating correlated systems. The proposed correlation between photoemission and thermoelectric power opens up a new possible approaches to research on the electronic structure of correlated actinide materials. This approach benefits from the interaction between fundamental and applied physics by relating the results obtained by two different methods, which previously have been found difficult to balance.

\begin{figure}[t!]
\begin{centering}
\includegraphics[width=0.47\textwidth]{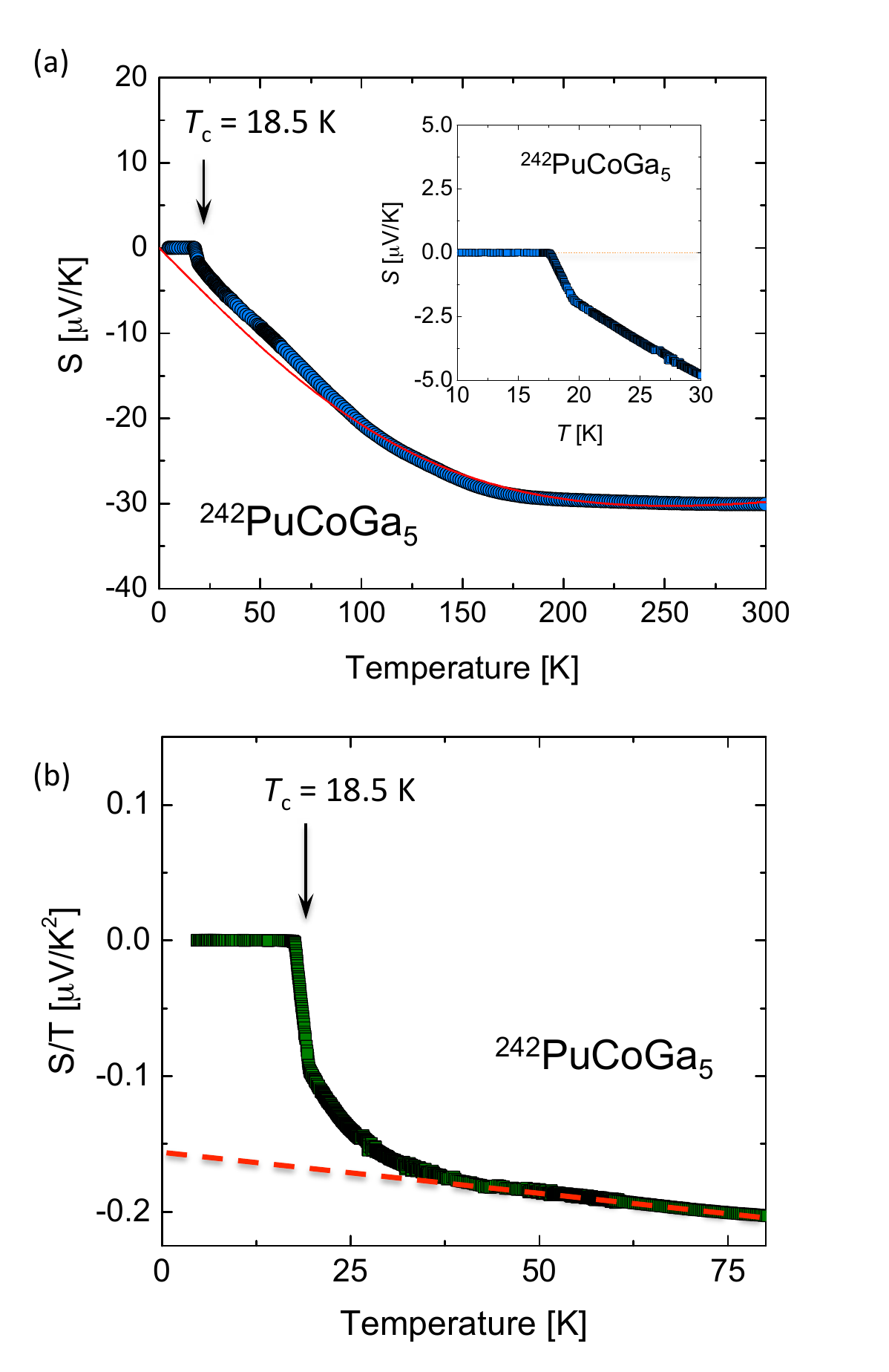}
\caption{(Color online) (a) The temperature dependence of the thermoelectric power of PuCoGa$_{5}$. A superconducting transition is clearly visible as a sharp transition to zero at 18.5~K (see the inset). In the normal state, the temperature dependence of $S(T)$ is characteristic of intermediate valence systems and can be fit by the model described in the text (see the red line). (b) The low-temperature variation of $S/T$. The dashed red line is an extrapolation of $S(T)/T$ to zero temperature. The arrows mark the superconducting phase transition at $T_{c}$ = 18.5 K.}\label{1}
\end{centering}
\end{figure}

Polycrystalline samples of PuCoGa$_{5}$ were prepared by arc melting stoichiometric amounts of the elemental components in a Zr-gettered argon atmosphere. In order to reduce a self-heating effect and related disorder associated with the decay of $^{239}$Pu isotope \cite{LANL} a much more stable, but less common, isotope of Pu element, namely $^{242}$Pu, has been used in these studies. The samples were examined by x-ray powder diffraction and the phase composition was determined by energy dispersive x-ray analysis with a Philips XL40 scanning electron microscope. The crystal structure was shown to be a tetragonal with the HoCoGa$_{5}$-type with lattice parameters similar to the those reported in literature.\cite{Pu115,jcg} The thermoelectric power was measured using a custom-made device in the temperature range of 2 - 300~K using HFC copper as a reference material. Due to the radioactivity of Pu, all operations were performed in a nitrogen inert atmosphere glove-box and a special encapsulation procedure was used in order to avoid any contamination risk.

Thermopower measurements have gained importance in recent years as a method of choice for investigating thermoelectric materials that are potential candidates for applications, such as spot cooling of electronic components, waste heat recovery systems and/or remote power generation in space stations and satellites (see Refs.\onlinecite{t1,t2} and references therein). In addition, the Seebeck coefficient is a sensitive probe of density of states relative to the Fermi level, it can therefore be used as a tool to characterize the electronic structure of materials, especially in the vicinity of the narrow gap or pseudo-gap. The temperature dependence of the thermoelectric power of PuCoGa$_{5}$ is shown in Fig.\ref{1}. In general, the overall shape and magnitude of the $S(T)$ curve is typical of strongly correlated intermediate valence systems (see Refs. \onlinecite{gott,kot,dk,z5,z6} and references therein). At room temperature the magnitude of the thermopower is strongly enhanced and has a value of about -30~$\mu$V/K. This value of $S$ is one order of magnitude larger than the values observed in simple metals such as copper or silver.\cite{bl} The strongly enhanced value of the Seebeck coefficient indicates that 5$f$ electrons participate in the bonding and are present at the Fermi energy. Furthermore, the negative sign indicates that carrier electrons might dominate thermal and electrical conduction, and is consistent with recent theoretical calculations.\cite{jxz} Below the superconducting transition at $T_{c}$ = 18.5 K, the Seebeck coefficient of PuCoGa$_{5}$ drops to zero (see Fig.\ref{1}) as expected for a bulk superconducting state\cite{bl} and is in agreement with other studies of this material.\cite{Pu115} It has been shown by Behnia {\it et~al.}\cite{behnia} that for many correlated materials, in the limit of zero temperature, the thermopower mirrors the specific heat per electron.\cite{ziman} By taking into account these two quantities the dimensionless coefficient $q$ can by identified as:\cite{behnia}

\begin{equation}
q=F\frac{S}{T}\gamma^{-1}\label{q}
\end{equation}

where $F$ is the Faraday constant ($F=eN_{A}$ = 96,485.34 C~mol$^{-1}$) and $\gamma$ is the low temperature specific heat ($C/T$). The dimensionless quantity $q$ corresponds to the density of carriers per formula unit for the case of a free electron gas with an energy independent relaxation time. This coefficients has been shown to be close to unity for many Ce base strongly correlated materials (1 $e^{-}/f.u.$). For higher (lower) density than 1 $e^{-}/f.u.$, the absolute value of $q$ is expected to be proportionally smaller (larger) than unity.\cite{behnia} As shown in Figure \ref{1}b, the extrapolated $S/T$ value for PuCoGa$_{5}$ is 0.16~$\mu$V/K$^{2}$. This value together with $\gamma \approx$~77~mJ/mol K$^{2}$ estimated from superconducting jump at $T_{c}$ (see Ref.~\onlinecite{Pu115}) gives $q \approx$ 0.2. This value of $q$ might suggest that all five 5$f$-electrons are delocalized in PuCoGa$_{5}$.

The curvilinear character of $S(T)$ is indicative that PuCoGa$_{5}$ has a complex electronic structure, especially in the vicinity of the Fermi energy. Indeed, photoemission measurements\cite{joyce} and theoretical calculations\cite{jxz} show a relatively complicated electronic structure in this compound, characterized by the presence of 5$f$-electrons close to the Fermi energy. In general, several contributions may effect the density of states within the Fermi window (2$k_{B}T$) such as a Kondo resonance or crystal field effects.\cite{TEP1} In the case of intermediate valence systems, the Kondo temperature is greater than the crystal field splitting and, therefore, the $f$-states for a Kondo impurity are characterized by single quasiparticle peak located in the close vicinity of the Fermi level.\cite{TEP1}

\begin{figure}[t!]
\begin{centering}
\includegraphics[width=0.47\textwidth]{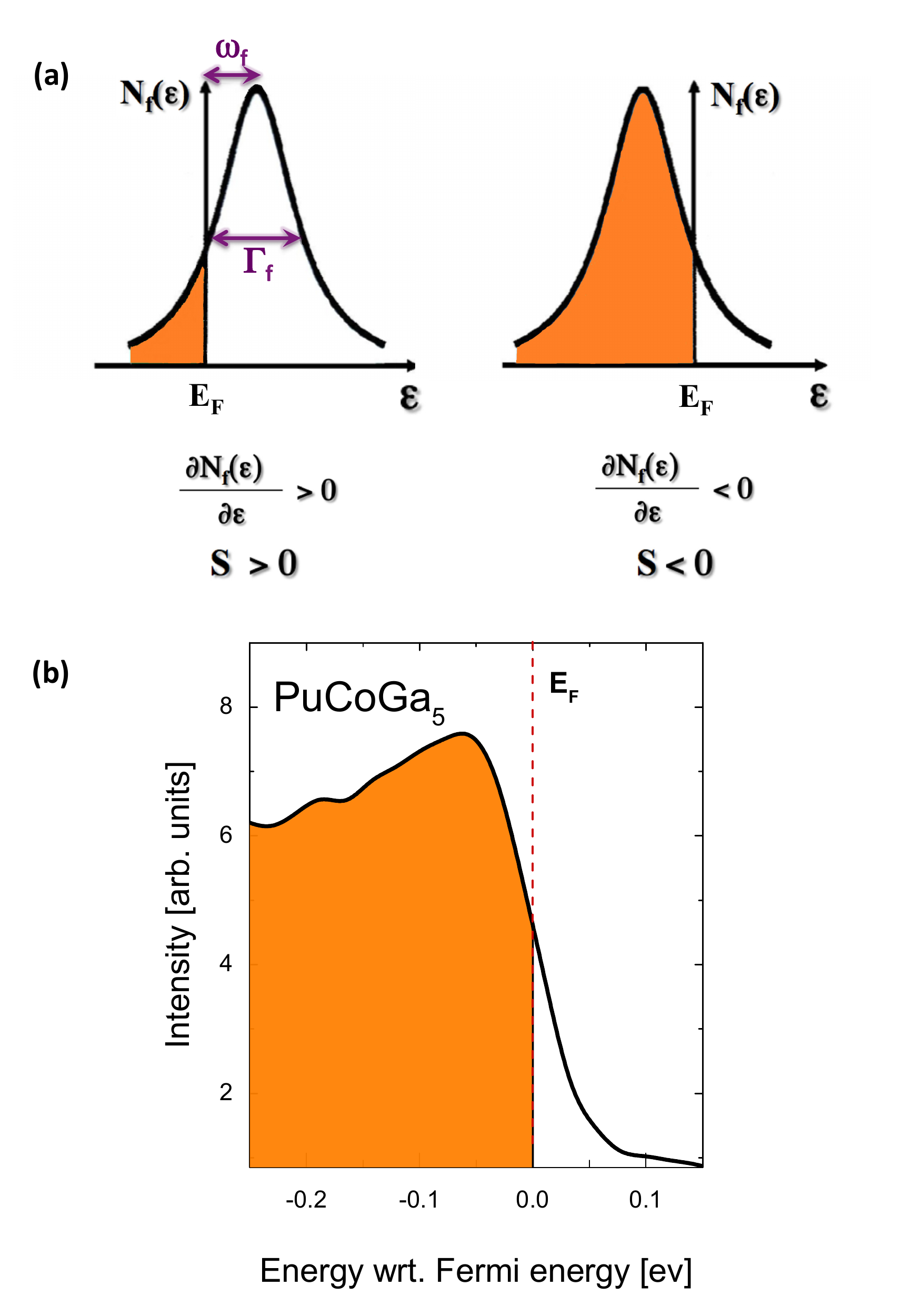}
\caption{(Color online) (a) The model of the density of states in the vicinity of the Fermi energy.\cite{gott} The Lorenzian density of states is characterized by width $\Gamma_{f}$ and its position with relation the the Fermi energy $\omega_{f}$ (see text). (b) The photoemission spectrum of PuCoGa$_{5}$ in the vicinity of Fermi energy (data taken from Ref.~\onlinecite{joyce})}.\label{2}
\end{centering}
\end{figure}

To describe the temperature variation of the Seebeck coefficient in such compounds, we use a phenomenological resonant level model\cite{gott} where the narrow $f$ quasiparticle band is assumed to have a Lorentzian form:

\begin{equation}
N_{f}(E)={1 \over \pi} \frac{\Gamma_{f}}{(E-\omega_{f})^{2}+\Gamma_{f}^{2}},
\end{equation}

in which $\omega_{f}$ represents the energy of the Lorentzian relative to the Fermi energy and $\Gamma_{f}$ is its width (see Fig.\ref{2}a). In this approach, the conduction electrons are assumed to scatter from the 5$f$ quasiparticles, so that the thermoelectric power may be approximated by a modified Mott formula:\cite{bando,kot}

\begin{equation}
S_{f}(T)=\frac{2}{3}\pi^{2}\frac{k_{B}}{|e|}\frac{\omega_{f}T}{T^{2}(\pi^{2}/3)+\omega_{f}^{2}+\Gamma_{f}^{2}},
\end{equation}

in agreement with the behavior of many strongly correlated materials.\cite{adamprb,kot,dk,bando,gott} For a normal metal, the low-temperature limit follows a linear $T$ behavior characteristic of a Fermi-Liquid. As shown by the solid lines in Fig.\ref{1}a, the above model provides a good description of the experimental data of PuCoGa$_{5}$ in the normal state. The data can be fit with the parameters $\omega_{f}$ and $\Gamma_{f}$, with values -7.4 and 35.1 meV, respectively. Figure \ref{2}b shows the photoemission spectra of PuCoGa$_{5}$ in the vicinity of Fermi energy, $E_{F}$ (data taken from Ref.~\onlinecite{joyce}). A sharp peak, originating from the 5$f$ states, is observed below $E_{F}$ pointing to itinerant nature of 5$f$-electrons. The spectrum can be fit by the parameters, $\omega_{f}$ and $\Gamma_{f}$, with values of -36 and 149 meV, respectively.

\begin{figure}[t!]
\begin{centering}
\includegraphics[width=0.47\textwidth]{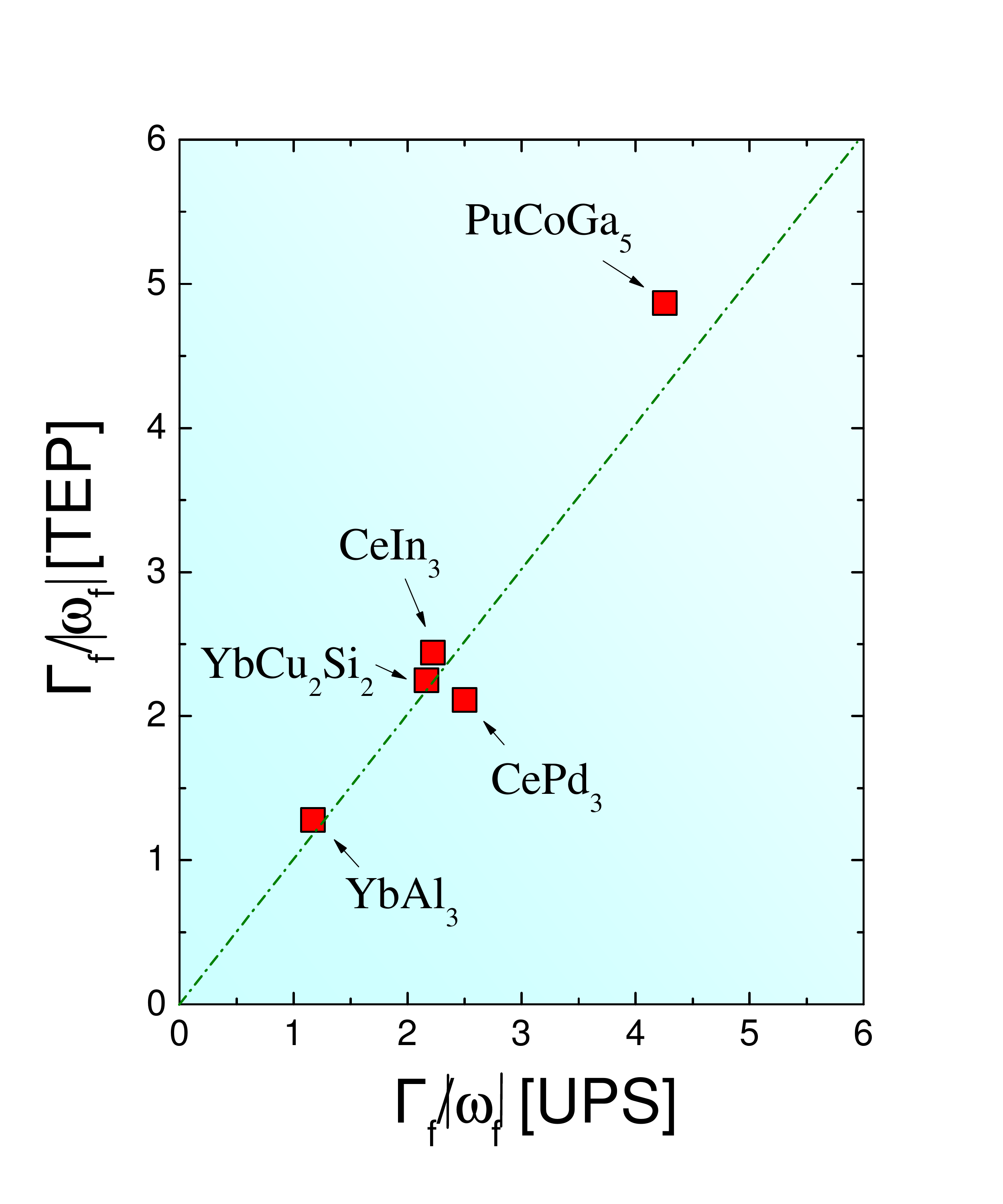}
\caption{(Color online) $\Gamma_{f}$/$\omega_{f}$ obtained from thermoelectric measurements (TEP) versus $\Gamma_{f}$/$\epsilon_{f}$ derived from photoemission studies (UPS) for various Ce-, Yb-, and Pu-based intermediate valence systems (see text). For the plot, data have been taken from: YbPd$_{3}$\cite{TEP_YbPd3}, CeIn$_{3}$\cite{TEP_Ce13}, CePd$_{3}$\cite{TEP_Ce13}, YbCu$_{2}$Si$_{2}$.\cite{TEP_Yb122}}\label{3}
\end{centering}
\end{figure}

Both the thermopower and the photoemission experiments can be interpreted in terms of a simple scaling framework, such as found in the Kondo model. In the Kondo impurity model in which the energy of the resonance is given by $\omega_f = k_B T_K$ and the width is given by $\Gamma_f = \pi { k_B T_K/N}$, where $N$ is the degeneracy of the lowest spin-orbit multiplet. This one-parameter scaling of the Kondo model results in the ratio ${\Gamma_f/ \omega_f}$ having the well-defined value ${\pi/N}$. For concentrated compounds, such as the Anderson Lattice Model, it has been proposed that two energy scales exist.\cite{Burdin} One scale represents a Kondo temperature at which the $f$ moments start to be screened. The second and smaller scale is the coherence temperature $T_0$, which characterizes the low-temperature Fermi-Liquid and below which the moments are fully-screened. These two scales were found to have the same exponential variations but have different pre-factors.\cite{Burdin} Further support for the existence of two scales differing only by a numerical factor can be found by adapting Nozieres arguments\cite{Nozieres1,Nozieres2} by assuming that the renormalized hybridization matrix element is the fundamental energy scale for the lattice and by including the spin-degeneracy $N$.\cite{R&L} In this case, one finds that the energy scale that characterizes the formation of the hybridized bands is given by $k_B T_K$ below which only one component of each spinor representing the magnetic moment at any site is randomized. The second scale becomes $k_B T_c$ which is approximately given by ${k_B T_K / N}$, at which all the $N$-components of the spinors are randomized leading to a non-magnetic ground state. If the ratio of the two scales is determined by the quasi-particle mass enhancement factor\cite{Burdin} $Z=1-\frac{\partial \Sigma}{\partial \omega}$, then the ratio $\Gamma_f / \omega_f$ obtained from both measurements will be invariant. Therefore, we hypothesize that the parameters that describe the coherent, low-temperature Fermi-Liquid-like variation of the thermopower should be related to the coherence temperature $T_0$, while the photoemission spectrum should be characterized by $T_K$. In this case, although the values of $\omega_f$ and $\Gamma_f$ may differ for these two measurements, their ratio is still well-defined. Consequently, in Figure \ref{3} we plot $\Gamma_{f}$/$\omega_{f}$ obtained from thermoelectric measurements (TEP) versus $\Gamma_{f}/\omega_{f}$ derived from photoemission studies (UPS) for PuCoGa$_{5}$. For comparison we have also included the thermoelectric and photoemission results derived for various 4$f$- and 5$f$-electron based intermediate valence systems.\cite{wyj} Furthermore, similar trends are also observed in other 4$f$- and 5$f$-based correlated materials that show itinerant behavior. The observed linearity supports our hypothesis and leads to the conclusion that the thermopower can be used as a probe of the density of states in this class of materials. It should be noted that the above model is very simple and neglects some interactions that may be important in crystalline materials. In more rigorous approaches, especially those in which the $f$-electrons are more localized, one should take the effect of crystal fields and their influence on the electronic band structure into account. In the Coqblin and Schrieffer (CS) model \cite{cob1,cob2}, the Kondo effect is described in the presence of the crystal field interactions which strongly influence the temperature variations of the Seebeck coefficient. Recently, Zlati$\acute{c}$ {\it et al.} extended the CS model by considering, besides the crystal field interactions, coherent hybridization between the $f$ and conduction electrons.\cite{z1,z2,z3,z4} The calculations were successful in describing the temperature dependencies of the thermoelectric power of various strongly correlated Ce- and Yb-based materials.\cite{z5,z6} Similar approach should also be applied to Pu-base correlated materials.\\

To summarize, we present for the first time the thermoelectric properties of plutonium superconductor PuCoGa$_{5}$. The compound is a correlated material with a superconducting transition temperature of $T_{c}$~=~18.5~K. We show that the Seebeck coefficient of PuCoGa$_{5}$ is strongly enhanced and its magnitude and temperature dependence is characteristic of intermediate valence systems in which 5$f$-states are present just below the Fermi level. All the results obtained point to the itinerant nature of the 5$f$-electrons and their importance for superconductivity in this material. In addition, the presence of valence fluctuations in PuCoGa$_{5}$ may be important as bosons which meditate the superconductivity.\cite{VF} The parameters which characterize the electronic structure that are obtained from the thermoelectric measurements are linearly related to the electronic structure parameters obtained from UPS. The approach adopted for PuCoGa$_{5}$ can be extended to other itinerant valence fluctuating correlated systems. The correlation between photoemission and thermoelectric power opens a new way to look at electronic structure of correlated nuclear materials. However, given the simplicity of the model, further studies are required to better understand the electronic structure of these materials. Also, it would be important to update the TEP/UPS scaling proposed here with more itinerant correlated compounds. We hope that our paper will stimulate such an attempt.

\begin{acknowledgments}

This work was supported by the Department of Energy, Office of Basic Energy Sciences, Materials Sciences, and Engineering Division and through grant DOE FG02-01ER45872. We are grateful to J. Rebizant for sample preparation and characterization. High purity Pu metal was made available through a loan agreement between Lawrence Livermore National Laboratory and ITU, in the frame of a collaboration involving LLNL, Los Alamos National Laboratory and the US Department of Energy.

\end{acknowledgments}


\begin{thebibliography}{99}
\bibitem{wyj} To the best of our knowledge we could not find more examples of correlated 4- and 5$f$-electron systems showing itinerant behavior for which both reliable thermoelectric and spectroscopic data exist.
\end{thebibliography}
\end{document}